\documentclass[letterpaper, paper,11pt]{AAS}		

\usepackage{bm}
\usepackage{amsmath, amssymb}
\usepackage{subfigure}
\usepackage{txfonts}
\usepackage[colorlinks=true, pdfstartview=FitV, linkcolor=black, citecolor= black, urlcolor= black]{hyperref}
\usepackage{overcite}
\usepackage{footnpag}			      	

\PaperNumber{25-254}

\begin{document}

\title{Communication  Constellation  Design of Minimum Number of Satellites with Continuous Coverage and Inter-Satellite Link}

\author{Soobin Jeon\thanks{PhD Candidate, Department of Astronomy, Yonsei University, 50 Yonsei-ro, Seodaemun-gu, 614A Science building, Seoul, Korea.},  
Sang-Young Park\thanks{Professor, Department of Astronomy, Yonsei University, 50 Yonsei-ro, Seodaemun-gu, 624 Science building, Seoul, Korea.}
}

\maketitle{}

\begin{abstract}
The recent advancement in research on distributed space systems that operate a large number of satellites as a single system urges the need for the investigation of satellite constellations. Communication constellations can be used to construct global or regional communication networks using inter-satellite and ground-to-satellite links. This study examines two challenges of communication constellations: continuous coverage and inter-satellite link connectivity. The bounded Voronoi diagram and APC decomposition are presented as continuous coverage analysis methods. For continuity analysis of the inter-satellite link, the relative motion between adjacent orbital planes is used to derive analytic solutions. The Walker–Delta constellation and common ground-track constellation design methods are introduced as examples to verify the analysis methods. The common ground-track constellations are classified into quasi-symmetric and optimal constellations. The optimal common ground-track constellation is optimized using the BILP algorithm. The simulation results compare the performance of the communication constellations according to various design methods. 
\end{abstract}

\section{Introduction}

In recent years, the feasibility of low Earth orbit communication satellites has attracted attention owing to the trend of small satellites. Examples include Space-X's Starlink, Eutelsat's OneWeb, and Amazon's Kuiper projects \cite{threeLEOconst}. Their orbits are designed to provide communication links across the Earth. A satellite constellation operates several satellites with distinct orbits in a single system. Compared with conventional geostationary satellites, low Earth orbit satellites have a smaller coverage size and shorter orbital periods, resulting in degraded spatial and temporal coverage performance. Therefore, low Earth orbit constellations operate several satellites to overcome this limitation while minimizing the number of satellites. If the coverage is continuous in the area of interest, the communication constellation can continuously provide communication links. In addition, when the inter-satellite links are connected, the link is not disconnected even by the orbital motion of the satellite and the Earth's rotation effect. In this study, the low Earth orbit communication constellation design problem was interpreted as the problem of achieving continuous coverage and inter-satellite links.

The  sections are presented in the following order: constellation design methods, coverage analysis methods, relative motion in adjacent orbital planes, simulation results, and conclusions. Constellation design methods describe the basis of Walker and common ground-track constellations. The Walker constellation first introduces the concept of a seed satellite and its design parameters, and defines the orbital elements. The pattern repetition period and duplicate allocation of orbital elements are explained as two important characteristics of the Walker-Delta constellation, and are expressed by the design parameters. The repeat ground-track orbit is the seed satellite orbit for a common ground-track constellation. The quasi-symmetric method configures satellites such that the spacing between adjacent satellites is almost equal. The BILP method optimizes the satellite configuration to satisfy the coverage requirements while minimizing the total number of satellites. The coverage analysis methods section first explains a useful concept: the geometry of the Earth’s coverage. Then, Voronoi tessellation and APC decomposition are described with the references. The relative motion in adjacent orbital planes derives the key formulae to analyze the bounds of the relative distance. The simulation results present the continuous coverage analyses for the Walker-Delta, quasi-symmetric, and BILP constellations, and the inter-satellite link continuity analyses for the three constellations. The last section summarizes and concludes the contents of this paper.

\section{Constellation Design Methods}

\subsection{Walker Constellation}
The Walker constellation is a geometric design method that symmetrically configures the orbits \cite{JGWalker1970,JGWalker1977}. The seed satellite determines the common orbital elements of the constellation. For the Walker constellation, the satellites are configured in circular orbits with the same semi-major axis ($a$), inclination ($i$), and argument of perigee ($\omega$) of the seed satellites. The two types of Walker constellations are classified according to their inclinations. The Walker-Star constellation is designed based on a polar orbit and achieves global coverage. On the other hand, the Walker-Delta constellation has an inclination under 90 degrees and covers the mid-latitude and equator regions. Therefore, this classification is relevant only to the latitude of the area of interest (AoI). 

\begin{figure}[b!]
	\centering\includegraphics[width=3.5in]{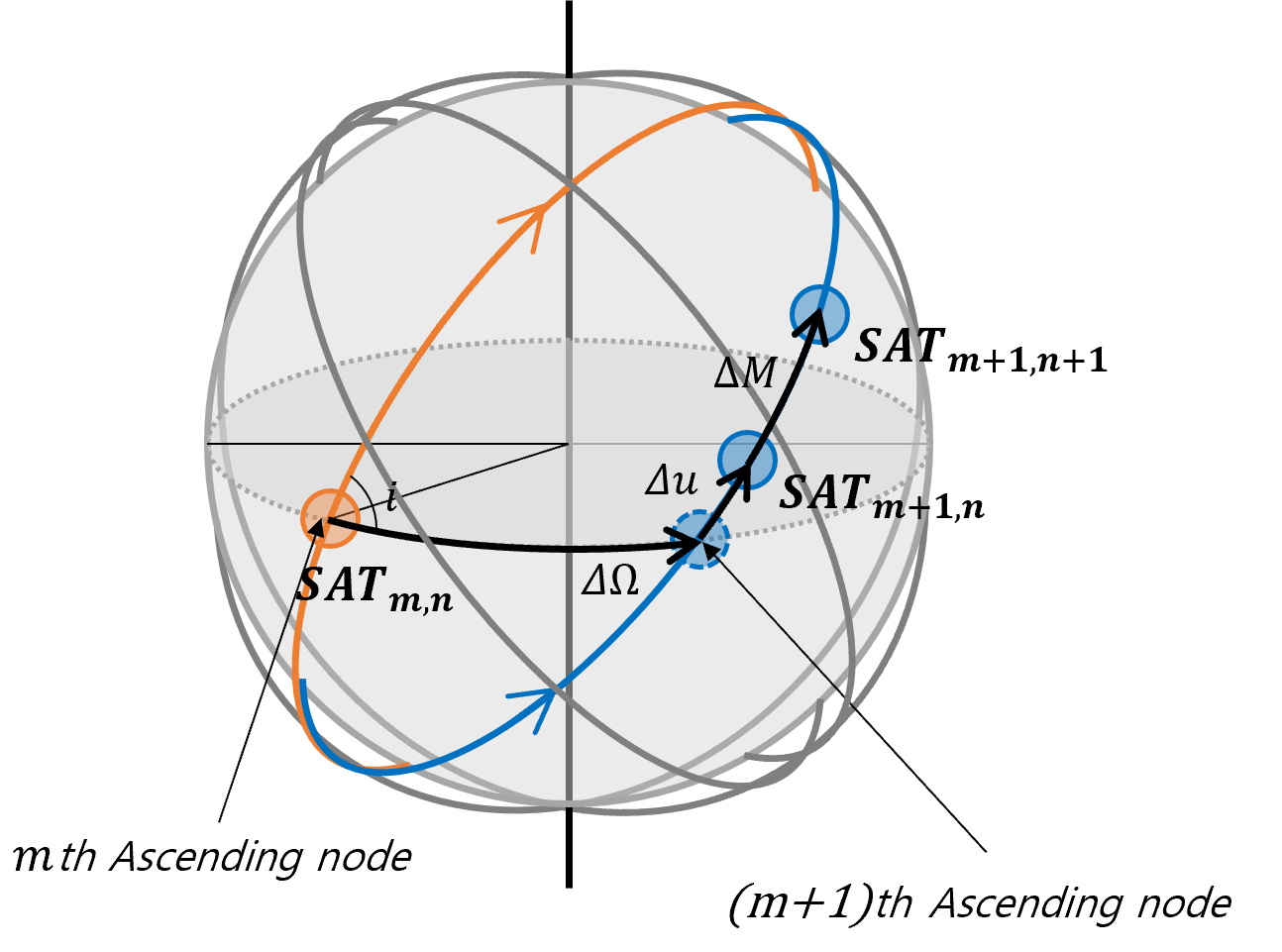}
	\caption{Geometric description of the orbital elements of the Walker-Delta constellation}
	\label{figWD}
\end{figure}

The three design parameters are the total number of satellites ($T$), number of orbital planes ($P$), and phasing parameter ($F\in \{0,1,..., P-1 \}$). The number of satellites per orbital plane ($S = T/P$) is an auxiliary parameter used to prevent confusion. The right ascension of ascending node (RAAN, $\Omega_{m}$) and mean anomaly ($M_{m,n}$) of the $n$th satellite in the $m$th orbital plane ($SAT_{m,n}$) are defined by Eqs.~\eqref{OmMWD1} and \eqref{OmMWD2}, as shown in Figure~\ref{figWD}.
\begin{equation}\label{OmMWD1}
    \Omega_{m} = \frac{360}{P} \cdot \left( m-1 \right) \deg
\end{equation}
\begin{equation}\label{OmMWD2}
    M_{m,n} = \frac{360}{T} \cdot F \cdot \left( m-1 \right) + \frac{360}{S} \cdot \left( n-1 \right) \deg
\end{equation}
where $m=1,2,...,P$ and $n=1,2,...,S$.

From the Eq.~\eqref{OmMWD1}, the RAAN is equally spaced by $\Delta \Omega = 360/P \deg$ in the range $\Omega_{m} \in \left[ 0, 360 \right]$ $\deg$. The intraplane spacing $\Delta M = 360/S \deg$ generates a symmetric location in the orbital plane in the range $M \in \left[ 0, 360 \right] \deg$ and is equivalent to the relative angular distance between the satellites in the orbital plane. The relative argument of the latitude ($\Delta u$) in the adjacent orbital planes is derived as $\Delta u = 360/T \cdot F \deg$ from Eq.~\eqref{OmMWD2}.

\subsubsection{Pattern Repetition Period}
The Walker-Delta pattern has geometric characteristics such that the intra- and inter-plane angular spacings are homogeneous. This geometric symmetry determines the pattern repetition period. The investigation of patterns in a geographical coordinate system enhances the understanding of the pattern repetition period. Reference ~\citenum{SJeonConst} derived the formulas for the pattern repetition period. The pattern unit (PU), which is introduced to understand the orbital configuration of the Walker-Delta constellation, is defined as follows \cite{JGWalker1970,JGWalker1977}:
\begin{equation}
    1 \text{PU} = \frac{360}{T} \deg
\end{equation}
The RAAN and mean anomaly in Eqs.~\eqref{OmMWD1} and \eqref{OmMWD2} are reorganized in PU as
\begin{equation}
    \Omega_{m} = S \cdot \left( m-1 \right) \text{PU}
\end{equation}
\begin{equation}
    M_{m,n} = F \cdot \left( m-1 \right) + P \cdot \left( n-1 \right) \text{PU}
\end{equation}
The time interval in which the mean anomaly increases for $F$ PU is the first pattern repetition period $t_{F}$ and defined as
\begin{equation}
    t_{F} = F \cdot \frac{360}{T} \cdot \frac{1}{\omega_{orb}} \text{sec}
\end{equation}
where $\omega_{orb}$ denotes the orbital angular speed.
Assuming the twobody motion, the RAAN and mean anomaly at $t_{F}$ are derived as
\begin{equation}\label{RAANWDtF}
    \Omega_{m}\left( t_{F} \right) = \Omega_{m} \left( t_{0} \right) = S \cdot \left( m-1 \right) \text{PU}
\end{equation}
\begin{equation}\label{MAWDtF}
\begin{aligned}
    M_{m,n}\left( t_{F} \right) &= M_{m,n} \left( t_{0} \right) + F = F \cdot m + P \cdot 
    \left( n-1 \right)\text{PU} \\
    &= M_{m+1,n} \left( t_{0} \right) \text{PU}
\end{aligned}
\end{equation}
where $t_{0}$ denotes the epoch time. 

From Eqs.~\eqref{RAANWDtF} and~\eqref{MAWDtF}, the mean anomaly of the $n$th satellite in $m$th orbital plane at $t_{F}$, $M_{m,n} \left( t_F \right)$, is the same as the one of the $n$th satellite in $(m+1)$th orbital plane at $t_{0}$, $M_{m+1,n} \left( t_{0} \right)$. Thus, the constellation pattern observed at $t_{F}$ appears as a pattern at $t_{0}$ that is shifted westward as $\Delta \Omega \deg$.

\begin{figure}[htb!]
	\centering\includegraphics[width=6in]{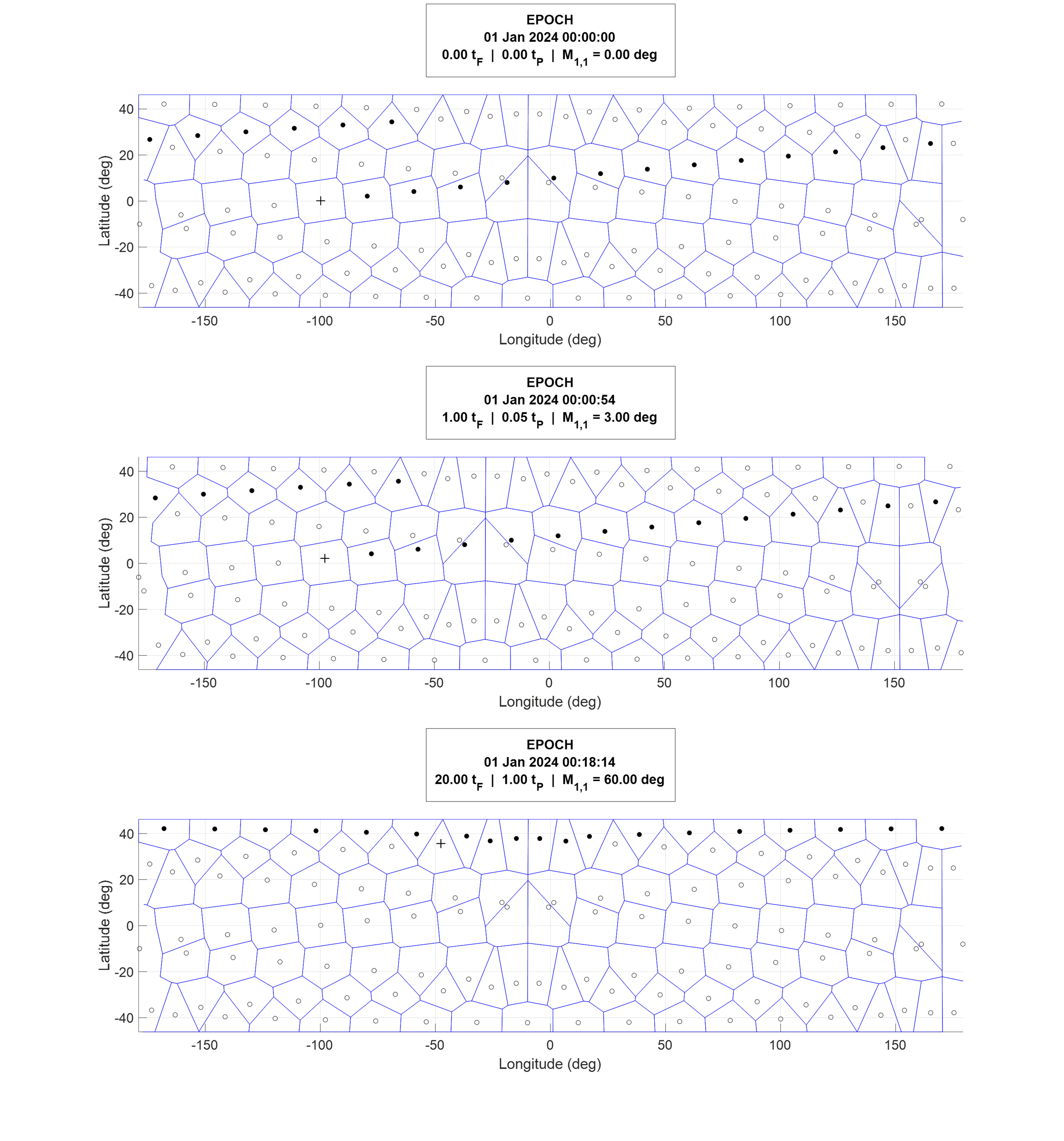}
	\caption{Walker-Delta patterns in geographic coordinate frame at epoch time, $t_{F}$, and $t_{P}$ ($+$: ${SAT}_{1,1}$, $\medbullet$: ${SAT}_{\textbf{M},1}$, $\medcirc$: the rest satellites)}
	\label{figRP}
\end{figure}

On the other hand, the mean anomaly propagates $P$ PU during $t_{P}$ and defines the second pattern repetition period as
\begin{equation}
    t_{P} = P \cdot \frac{360}{T} \cdot \frac{1}{\omega_{orb}} \text{sec}
\end{equation}
The RAAN and the mean anomaly at $t_{P}$ are derived as
\begin{equation}\label{RAANWDtP}
    \Omega_{m} \left( t_{P} \right) = \Omega_{m} \left( t_{0} \right) = S \cdot \left( m-1 \right) \text{PU}
\end{equation}
\begin{equation}\label{MAWDtP}
\begin{aligned}
    M_{m,n} \left( t_{P} \right) &= F \cdot \left( m-1 \right) + P \cdot n \text{PU} \\
    &= M_{m,n+1} \left( t_{0} \right)
\end{aligned}
\end{equation}
Let us define the set of satellites in the $m$th orbital plane as $\textbf{N} = \left\{n \mid n = 1,2,...,S\right\}$. Then, Eqs.~\eqref{RAANWDtP} and~\eqref{MAWDtP} derive 
\begin{equation}
    M_{m,\textbf{N}} \left( t_{P} \right) = M_{m,\textbf{N}} \left( t_{0} \right).
\end{equation}
Therefore, the constellation pattern at $t_{P}$ appears identical to the one at $t_{0}$.

Figure~\ref{figRP} shows the Walker-Delta constellation pattern for $i$: $T/P(S)/F$ = $42$: $120/20(6)/1$ in the geographic coordinate frame. The marks are subsatellite points; $+$, $\medbullet$, and $\medcirc$ represent ${SAT}_{1,1}$, ${SAT}_{\textbf{M},1}$, and the remaining satellites, respectively, where $\textbf{M} = \left\{m | m = 1,2,...,P \right\}$ is the set of orbital plane numbers. The blue line represents the bounded Voronoi diagram for the mid-latitude and equatorial regions, which will be described in the next section. ${SAT}_{1,1}$ is barely moved in the middle panel compared to the one in the top panel because $t_{F}$ is only 54 seconds. However, the blue diagrams show that the entire constellation moves westward for $\Delta \Omega = 18.00 \deg$. The bottom panel shows the pattern at $t_{P}$, which is 18 min and 14 s. The position of each satellite in the bottom panel is propagated for $t_{P}$ from the top panel; however, the patterns of the top and bottom panels are the same.

\subsubsection{Duplicate Allocation of Orbital Elements}
The Walker-Delta design method allocates the six unique orbital elements to each satellite, and the six orbital elements correspond to the unique orbital state of the six positions and velocity elements. This suggests the possibility of duplicate satellite positions and implies the collision between satellites. The conditions of the duplicate orbital elements of the Walker-Delta constellation are investigated in reference~\citenum{SJeonConst} as
\begin{equation}\label{dup}\begin{cases}
    \Omega_{m',n'} - \Omega_{m,n} = 180 \deg \\
    M_{m',n'} - M_{m,n} = 180 \deg.
\end{cases}
\end{equation}
Equation~\eqref{dup} can be expressed by the Walker-Delta design parameters as follows:
\begin{equation}\label{dupWD}
    \begin{cases}
        m' = 1 + \frac{P}{2} \\
        n' = mod \left( 1 + \frac{S-F}{2} ,S \right) + S \cdot \delta \left( mod \left( 1 + \frac{S-F}{2} ,S \right),0 \right) .
    \end{cases}
\end{equation}
where $mod \left(x,y \right)$ denotes the modulo operation, which returns the remainder when $x$ is divided by $y$, and $\delta \left(x,y\right)$ represents the Kronecker delta function, which equals $1$ if $x=y$ and $0$ otherwise. 
Therefore, the Walker-Delta design parameters that accommodate Eq.~\eqref{dupWD} must be avoided in the design procedure.

\subsection{Common Ground-track Constellation}

\subsubsection{Repeat Ground-track Orbit}
The repeat ground-track (RGT) orbit is an orbit that traces the same ground-track within a specific time interval. The two main parameters that determine the RGT orbit are the number of revolutions to repeat ($N_{P}$) and the number of days to repeat ($N_{D}$)\cite{Vallado}. For example, if the number of revolutions to repeat is $14$ and the number of days to repeat is $1$, the ground-track crosses (ascends or descends) the equator 14 times in one day. Thus, the period ratio ($\nu$), which is the RGT design parameter, is formulated as
\begin{equation}\label{tauNpNd}
    \nu = {N_{P}}/{N_{D}}
\end{equation}
The period ratio can also be described by the satellite nodal period ($T_{S}$) and the nodal period of Greenwich ($T_{G}$). The change in the orbital elements due to $J_{2}$ effect induces changes in $T_{S}$ and $T_{G}$ as
\begin{equation}\label{TS}
    T_{S} = \frac{2\pi}{\dot{\omega} + \dot{M}}
\end{equation}
\begin{equation}\label{TG}
    T_{G} = \frac{2\pi}{\omega_{E} - \dot{\Omega}}
\end{equation}
where $\omega_{E}$ is Earth's rotation speed. 

The orbital elements are formulated using Eqs.~\eqref{dotom}, \eqref{dotM}, and \eqref{dotOm}:
\begin{equation}\label{dotom}
    \dot{\omega} = \frac{3}{2}J_{2}\frac{R_{E}}{p}^{2} \sqrt{\frac{\mu_{E}}{a^{3}}} \left( 2 - \frac{5}{2}\sin{i}^{2} \right)
\end{equation}
\begin{equation}\label{dotM}
    \dot{M} = \sqrt{\frac{\mu_{E}}{a^{3}}} \left( 
1 - \frac{3}{2}J_{2}\left( \frac{R_{E}}{p} \right)^{2} \sqrt{1-e^2} \left( \frac{3}{2}\sin{i}^{2} -1 \right) \right)
\end{equation}
\begin{equation}\label{dotOm}
    \dot{\Omega} = -\frac{3}{2}J_{2} \left( \frac{R_{E}}{p} \right)^{2} \sqrt{\frac{\mu_{E}}{p}^{2}} \cos{i}
\end{equation}
Based on the definition of period ratio, Eqs.~\eqref{TS} and \eqref{TG} derive Eq.~\eqref{tauNpNd} with respect to the orbital elements as:
\begin{equation}\label{tauTGTS}
    \nu = \frac{N_{P}}{N_{D}} = \frac{T_{G}}{T_{S}} = \frac{\dot{\omega} + \dot{M}}{\omega_{E} - \dot{\Omega}}
\end{equation}

From Eqs.~\eqref{dotom},~\eqref{dotM}, and~\eqref{dotOm}, the orbital elements have arguments as $\dot{\omega} = \dot{\omega}\left( a,i,e \right)$, $\dot{M} = \dot{M} \left( a,i,e \right)$, and $\dot{\Omega} = \dot{\Omega} \left( a,i,e \right)$. It organizes the arguments of $T_{S}$, $T_{G}$, and $\nu$ as:
\begin{equation}\label{TSArg}
    T_{S} = T_{S} \left( a, e, i \right)
\end{equation}
\begin{equation}\label{TGArg}
    T_{G} = T_{G} \left( a, e, i \right)
\end{equation}
\begin{equation}\label{tauArg}
    \nu = \nu \left( a, e, i \right)
\end{equation}

From this, the RGT orbital design algorithm can be derived. Given the specific $\nu$, inclination ($i$), and eccentricity ($e$), the algorithm calculates the corresponding semi-major axis (a). Equation~\eqref{tauNpNd} implies that $\nu$ determines the number of revolutions per period and suggests that $\nu$ is related to the orbital period and semi-major axis. Consequently, when $e$ and $i$ are specified, one unique semi-major axis is derived from $\nu$ and vice versa. For example, if the set of RGT orbital elements is given as $\left( \nu , i, e \right) = \left( 14, 42 \deg, 0 \right)$, then $a$ is $7201.90 km$. As a result, the RGT orbital elements can be described in two different ways, such as $\left( a, i, e\right) = \left( 7201.90 km, 42 \deg, 0 \right)$, and then $\nu$ is uniquely determined as $14$. 

\subsubsection{Common Ground-track Constellation}
The RGT orbit that is designed according to Eqs.~\eqref{tauTGTS} contains a set of $\left( \nu, i, e \right)$ or $\left( a, i, e \right)$ with an arbitrary set of $\left( \omega, \Omega, M \right)$. From here on, the orbital elements of the RGT orbit are denoted as $\left( \nu, i, e \right)$ except for specific purposes. This implies that a numerous number of satellites can trace the same ground-track and introduces the concept of a common ground-track (CGT) constellation. The CGT constellation is designed following the three procedures below\cite{Flower,HLee2020}: \\
(1) Calculate the semi-major axis ($a$) of the seed satellite from $\left( \nu, i, e \right)$ in Eq.~\eqref{tauTGTS}.\\
(2) Choose an arbitrary $\omega$ so that all satellites have the same $\left( \nu, i, e, \omega \right)$. \\
(3) Given the CGT constellation's total number of satellites ($T$), the $k$th satellite's RAAN and mean anomaly $\left( \Omega_{k}, M_{k} \right)$ satisfy Eq.~\eqref{CGTomM}
\begin{equation}\label{CGTomM} 
    N_{P} \Omega_{k} + N_{D} M_{k} = constant \mod{2\pi}
\end{equation}
where $k=1,...,T$.

The CGT constellation design problem is concluded as the configuration method of $\left( \Omega_{k}, M_{k} \right)$ following procedure (3). The reference ~\citenum{HLee2020} introduces two methods, quasi-symmetric and binary integer programming (BILP). 

\subsubsection{Quasi-symmetric Method}

The simulation time ($T_{sim}$) is discretized by the step size $t_{step}$ and is assumed to be an integer multiple of the repetition period. Thus, this can be formulated as $T_{sim} = L \cdot t_{step}$. The continuous time variable $t\in \left[0, T_{sim} \right]$ is converted into the discretized time variable $\tau \in \left\{ 0,1,...,L-1 \right\}$. The configuration of the satellites in the CGT constellation can be expressed as time-shifted seed satellites because all satellites have the same ground-track traces. Therefore, the constellation pattern vector $x\left[\tau\right]$ can be defined as follows:
\begin{equation}\label{constPV}
    x\left[\tau \right] \triangleq \begin{cases}
            1 & \text{if $\tau = \tau_{k}$}, \\
            0 & \text{otherwise}. \\
        \end{cases}
\end{equation}
where $\tau_{k}$ is the temporal location of the $k$th satellite. 

In the discretized time domain, the constellation pattern vector has length $L$. If the total number of constellations is $T$, then the spacing constant $\xi$ is defined as follows:
\begin{equation}\label{eta}
    \xi \triangleq \frac{L}{T}.
\end{equation}

If $\xi$ is an integer, $\tau_{k}$ is equally spaced and the satellites are configured symmetrically. In contrast, if $T$ is not a divisor of $L$, then $\xi$ is not an integer that makes the index $\tau_{k}$ a rational number. In this case, only a quasi-symmetric configuration is possible. The formulation for both symmetric and quasi-symmetric constellation pattern vectors $\overline{x} \left[ \tau \right]$ is defined as
\begin{equation}
    \overline{x} \left[ \tau \right] \triangleq \sum_{k=1}^{T} \delta \left[ nint \left( \tau - \xi \left( k-1 \right) \right) \right] 
\end{equation}
where $nint$ denotes the nearest integer function. 

\subsubsection{Binary Integer Linear Programming Method}
Because the domain of the constellation pattern vector is defined as binary, the BILP method is introduced as an optimization algorithm. The BILP algorithm is a variant of the linear programming method that optimizes a linearized objective function with constraints and boundaries. The problem statement of the linear programming can be generalized as\cite{IP}
\begin{equation}\label{lpProb}
    \min_{x}{\textbf{\textit{c}}^{T} \textbf{\textit{x}}} \text{ subject to} \begin{cases}
            \textbf{A}\cdot\textbf{\textit{x}} \leq \textbf{\textit{b}} \\
            \textbf{A}_{eq} \cdot \textbf{\textit{x}} = \textbf{\textit{b}}_{eq}
    \end{cases}
\end{equation}
where $\textbf{\textit{x}}$ is the decision variable of length $L$, $\textbf{\textit{c}}^{T}\textbf{\textit{x}}$ is the objective function, $\textbf{A}\in\mathbb{R}^{Q \times L}$ and $\textbf{\textit{b}}\in\mathbb{R}^{Q}$ are the matrix and vector that constitutes the $Q$ numbers of inequality constraints, $\textbf{A}_{eq} \in \mathbb{R} ^{R \times L}$ and $\textbf{\textit{b}} \in \mathbb{R} ^{R}$ are the matrix and vector of $R$ numbers of equality constraints. 

In addition to Eq.~\eqref{lpProb}, the domain of the decision variable $\textbf{\textit{x}}$ must be defined. The domain sets $\mathbb{R}_{\geq0}$, $\mathbb{Z}_{\geq0}$, and $\mathbb{Z}_{2}$ induce the pure real and binary integer linear programming. It is also possible for mixed-integer linear programming to include both real and integer decision variables. Thus, the BILP has the same problem statement but constrains the domain of the decision variables as
\begin{equation}
    \textbf{\textit{x}} \in \mathbb{Z}_{2} ^{L}
\end{equation}

The objective of optimization is to obtain a constellation pattern vector that minimizes the number of satellites while satisfying the coverage requirement. Because the summation of $x\left[ \tau \right]$ is equal to $T$ by the definition of the constellation pattern vector in Eq.~\eqref{constPV}, the problem is formulated as
\begin{equation}\label{probBILP}
    \min_{x}{\textbf{\textit{1}}}^{T} \textbf{\textit{x}} \text{ subject to} \begin{cases}
        \textbf{V}_{0,j} \textbf{\textit{x}} \geq \textbf{\textit{f}}_{j}, & \forall j \in \mathcal{J} \\
        \textbf{\textit{x}} \in \mathbb{Z}_{2}^{L}
    \end{cases}
\end{equation}
where $j$ is the index for the $j$th grid, $\mathcal{J}$ is the set of grids in the area of interest, and $\mathbb{Z}_{2}$ is the binary integer number set. The matrix $\textbf{V}_{0,j} \in \mathbb{Z}_{2}^{L \times L}$ is a seed-satellite access profile circulant matrix, which is addressed in detail in the next section. 

\section{Coverage Analysis Methods}
\subsection{Geometry of Earth Coverage}

\begin{figure}[t!]
	\centering\includegraphics[width=6in]{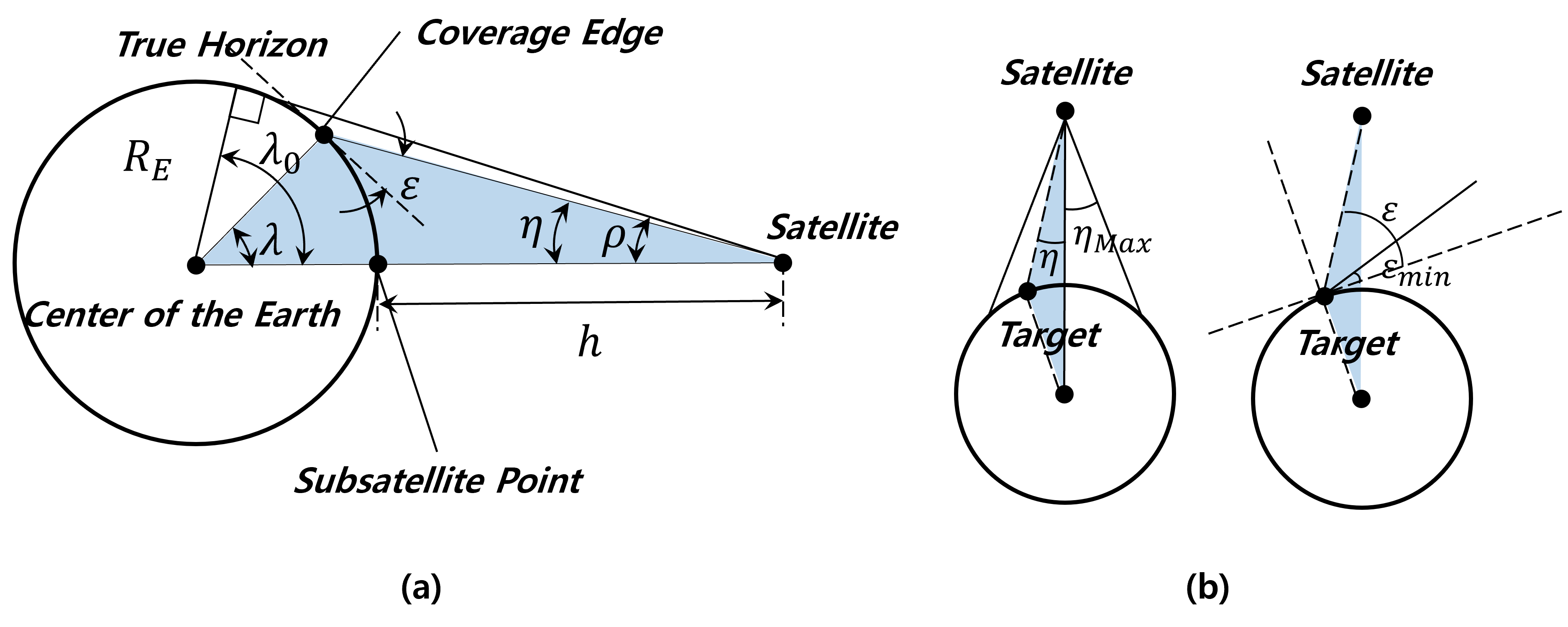}
	\caption{(a) Geometric relationships between the satellite, coverage edge, and center of the Earth and (b) geometric description of coverage}
	\label{figGeoCover}
\end{figure}

Figure~\ref{figGeoCover}a shows the geometric relationships between the satellite, coverage edge, and center of the Earth and is advantageous for coverage analysis. The true horizon is tangential from the satellite to the Earth's surface. The angle between the true horizon and the subsatellite point (SSP) is called the maximum Earth central angle (ECA, $\lambda_{0}$) or the angular radius of the Earth ($\rho$) when measured from the center of the Earth and the satellite, respectively. When the satellite is located at an altitude of h km, the angular radius of the Earth is determined using Eq.~\eqref{eqSinRho}:
\begin{equation}\label{eqSinRho}
    \sin{\rho} = \frac{R_{E}}{R_{E}+h}
\end{equation}
where $R_{E}$ is the Earth's radius.

Usually, the payload's coverage determines the coverage performance of a single satellite that constitutes the constellation. The Earth central angle ($\lambda$) measures the size of payload coverage ($\eta$) on the Earth’s surface and is defined as the angular distance between the SSP and the coverage edge. The coverage edge is the rim of the satellite coverage on the Earth. The elevation angle ($\varepsilon$) is measured at the coverage edge from the local horizontal to the satellite. 

The coverage can be approached from two perspectives: the satellite and the target area (Figure~\ref{figGeoCover}b). When considering the satellite perspective, it is important to evaluate payload's specifications. According to the definition of the coverage, the nadir angle ($\eta=\eta(t)$) should be smaller than the payload beam coverage ($\eta_{Max}$). From the target point perspective, it is considered to be covered when the elevation angle of the satellite ($\varepsilon = \varepsilon(t)$) is greater than the target point's minimum elevation angle ($\varepsilon_{min}$). 

The trigonometry of the blue shaded triangle in Figure~\ref{figGeoCover}a yields the formulae for $\eta$, $\varepsilon$, and $\lambda$ as Eqs.~\eqref{eqCosEps} and~\eqref{eqLambda}.
\begin{equation}\label{eqCosEps}
    \cos{\varepsilon} = \sin{\eta} / \sin{\rho}
\end{equation} 
\begin{equation}\label{eqLambda}
    \lambda = 90 \deg - \eta - \varepsilon
\end{equation}
For communication constellations, constraints are imposed on both the payload specification and the elevation angle of the target point. Therefore, the trigonometry of Eqs.~\eqref{eqCosEps} and~\eqref{eqLambda} is crucial, as it prevents redundant computational complexity. For example, suppose that the beam coverage of the spacecraft ($\eta_{Max}$) is $45 \deg$, and the ground station has a minimum elevation angle ($\varepsilon_{min}$) of $30 \deg$. If the satellite's altitude is $1,200$ km, the angular radius of Earth ($\rho$) is $57.31 \deg$. The Eq.~\eqref{eqCosEps} immediately converts the payload's coverage to the elevation angle as
\begin{equation}
    \overline{\varepsilon} = \arccos{ \left( \sin{\eta_{Max}} / \sin{\rho} \right) } = 32.84 \deg 
\end{equation}
where $\overline{\varepsilon}$ is the elevation angle corresponding to $\eta_{Max}$ and does not have a physical meaning. 
The ECA ($\overline{\lambda}$) is calculated using Eq.~\eqref{eqLambda} as
\begin{equation}
    \overline{\lambda} = 90 - \eta_{Max} - \overline{\varepsilon} = 12.16 \deg
\end{equation}
In the same manner, the minimum elevation angle ($\varepsilon_{min}$) yields its corresponding parameters as $\Tilde{\eta} = 46.79 \deg$ and $\Tilde{\lambda} = 13.21 \deg$. The ECA is the visualized size of the coverage on the Earth's surface; the smaller the ECA, the more degraded the coverage performance becomes. Therefore, a simulation with only $\overline{\lambda}$, $\eta_{Max}$, or $\overline{\varepsilon}$ is enough to analyze if the constellation satisfies the coverage requirement. Since the coverage analyses with $\overline{\lambda}$, $\eta_{Max}$, or $\overline{\varepsilon}$ show the same results but are conducted from different perspectives, the simulation with only one of the three constraints reduces the computational cost.

\subsection{Voronoi Tessellation}

The problem of a continuous coverage constellation can be reduced to obtaining the circumradius of three adjacent points. For a set of discretized points, the circumradius of three adjacent points can be defined. When the circumradius is smaller than a specified value, the distance from any point within the region is shorter than the specified value. References~\citenum{JGWalker1970} and~\citenum{JGWalker1977} introduced the satellite triad method as a coverage analysis technique for Walker constellations. The research subject of the references was a constellation of fewer than 20 satellites. This study utilizes the Delaunay triangulation method, which was first suggested in ~\citenum{SJeonConst}, to generalize the number of satellites in the constellation. 

Delaunay triangulation is a computational geometry method that subdivides discretized points into triangles\cite{BorisVD}. This algorithm defines the Delaunay criterion for constructing Delaunay conformant triangles that do not contain other points inside the circumcircles. The Voronoi diagram (VD) is a dual graph of the Delaunay triangle (DT) and is drawn by connecting the circumcenters of the Delaunay triangles. Voronoi tessellation refers to the tiling of a plane or sphere using Voronoi diagrams. If the tiled region is a restricted closed area on the sphere, the Voronoi diagrams have boundaries cut by the region and are called bounded Voronoi diagrams (BVD)~\cite{GDaiVD,SJeonConst}. The constellation coverage problem can be formulated as a Voronoi tessellation problem. The subsatellite points on the Earth's surface are discretized points on a three-dimensional spherical surface. The area of interest is not the entire globe, and the Voronoi diagram is bounded within the target region. The spherical Delaunay triangles and spherical bounded Voronoi diagrams derive the solution; however, the word `spherical' is omitted for brevity from here on. 

\begin{figure}[t!]
	\centering\includegraphics[width=6in]{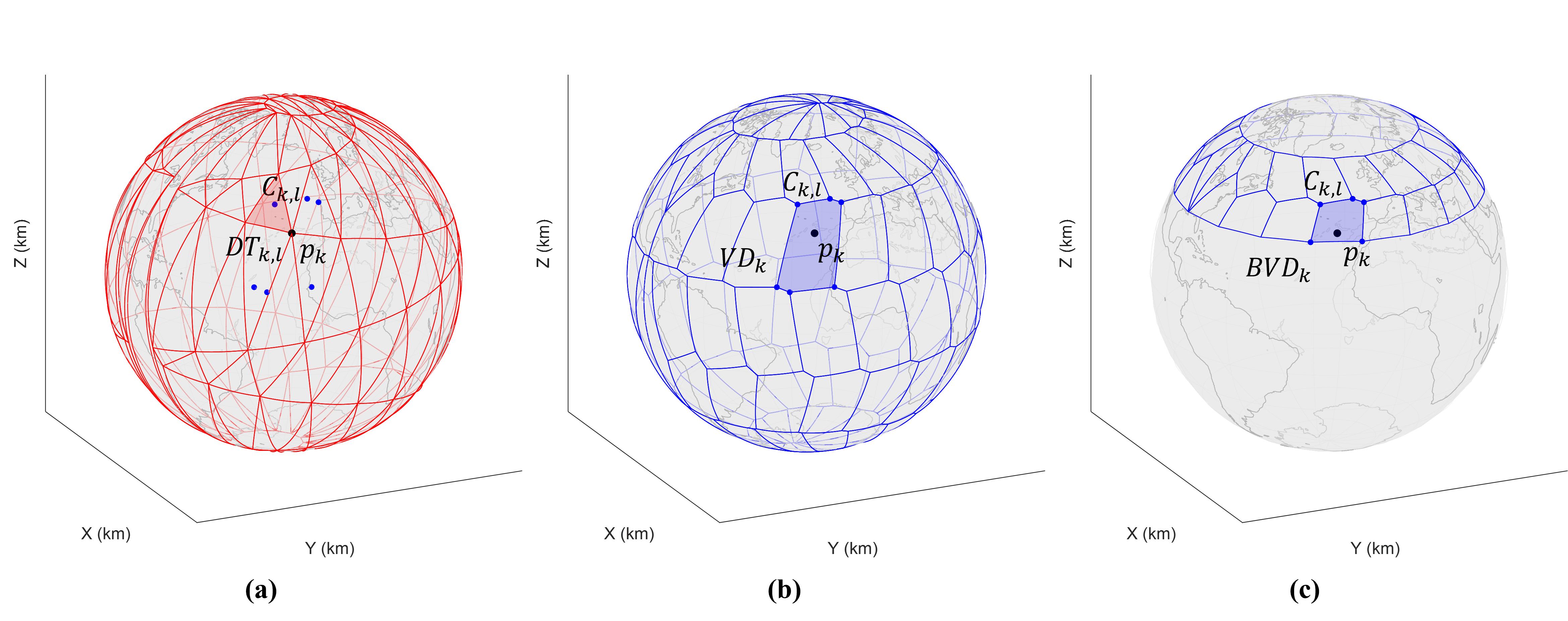}
	\caption{Example: (a) Delaunay triangle, (b) Voronoi diagram, and (c) bounded Voronoi diagram}
	\label{figDTVDBVD}
\end{figure}

Figure~\ref{figDTVDBVD} depicts examples of the Delaunay triangle, Voronoi diagram, and bounded Voronoi diagram. In Figure~\ref{figDTVDBVD}a, the  six DTs contain $p_{k}$ as their vertices for an arbitrary subsatellite point $p_{k}$, where $k = 1,2,...,T$. ${DT}_{k,l}$ are depicted as red triangles, and the circumcenters $C_{k,l}$ are blue dots, where $l=1,2,...,N_{k}$. $N_{k}$ is the number of triangles that have vertices $p_{k}$ and can be different for each $p_{k}$. Any subsatellite point possesses only one VD, and $VD_{k}$ is drawn by connecting the blue dots to the circumcenters $\textbf{C}_{k} = \left\{C_{k,l} \mid l = 1,2,...,N_{k} \right\}$, as shown in Figure~\ref{figDTVDBVD}b. 

The Voronoi diagram in Figure~\ref{figDTVDBVD}b can be used to solve the global coverage problem. However, for the regional coverage problem, the Voronoi diagram must be bounded within the area of interest. In particular, for the regional continuous coverage problem, the Voronoi diagram bounded within the latitude range of the AoI as a circular band can efficiently analyzes the continuity of the constellation\cite{SJeonConst}. Considering the northern area of interest, the bounded Voronoi diagram appears in Figure~\ref{figDTVDBVD}c. Then, $p_{k}$ has ${BVD}_{k}$ with different vertices $\overline{\textbf{C}}_{k} = \left\{ \overline{C}_{k,\overline{l}} \mid \overline{l} = 1,2,...,\overline{N}_{k} \right\}$. 

The angular distance $\theta_{k,\overline{l}}$ is defined as the angular distance between the subsatellite point $p_{k}$ and the vertices $\overline{C}_{k,\overline{l}}$. Then, the maximum distance ${BVD}_{k}$ of the $k$th satellite is expressed as:
\begin{equation}
    \theta_{max,k} = \max_{\overline{l}}{\theta_{k,\overline{l}}}
\end{equation}
Consequently, the maximum angular distance of the entire constellation is obtained as.
\begin{equation}
    \theta_{max} = \max_{k}{\theta_{max,k}}
\end{equation}
Assuming a homogeneous constellation, the coverage performances of all satellites are the same and can be expressed as $\lambda^{*}$. Then, the continuous coverage problem statement is defined as
\begin{equation}
    \theta_{max} \leq \lambda^{*}.
\end{equation}
For the global coverage problem, the maximum angular distance should be defined from $\theta_{k,l}$ and $C_{k,l}$, and the remaining procedures are the same.

\subsection{APC Decomposition}
The reference~\citenum{HLee2020} developed the APC decomposition based on the circular convolution phenomenon between a seed satellite's access profile, a constellation pattern vector, and a coverage timeline. The access profile between the $k$th satellite and the $j$th target point ($\textbf{\textit{v}}_{k,j}$) defines its elements as follows: 
\begin{equation}\label{vkj}
    v_{k,j} \left[ \tau \right] \triangleq \begin{cases}
        1 & \text{if $\varepsilon_{k,j} \left[ \tau \right] \geq \varepsilon_{k,j,min} \left[ \tau \right]$ } \\
        0 & \text{otherwise}
    \end{cases}
\end{equation}
where $\tau$ is the discretized time variable and $\mathcal{J}$ is the set of target points. 

The coverage timeline of constellation $\textbf{\textit{b}}_{j}$ is derived as the summation of all access profiles as follows:
\begin{equation}\label{bj1}
    b_{j} \left[ \tau \right] = \sum_{k=1}^{T} {v_{k,j}} \left[ \tau \right]
\end{equation}
The circular convolution phenomenon expresses the coverage timeline $\textbf{\textit{b}}_{j}$ with respect to the seed satellite access profile $\textbf{\textit{v}}_{0,j}$ and the coverage pattern vector $\textbf{\textit{x}}$ as
\begin{equation}\label{bj2}
    b_{j} \left[ \tau \right] = v_{0,j} \left[ \tau \right] \circledast x \left[ \tau \right]
\end{equation}
where $\textbf{\textit{x}}$ is defined in Eq.~\eqref{CGTomM} and $\circledast$ denotes the circular convolution operator.

This circular convolution operation can be described in a linearized form as follows:
\begin{equation}\label{bj3}
    \textbf{\textit{b}}_{j} = \textbf{V}_{0,j} \textbf{\textit{x}}
\end{equation}
where $\textbf{V}_{0,j}$ is the matrix in Eq.~\eqref{probBILP}. 

In summary, the coverage timeline of the entire constellation is obtained from the circular convolution of the seed satellite access profile and the constellation pattern vector. This concept of APC decomposition reduces the optimal CGT constellation design problem to a constellation pattern vector optimization problem. 

\section{Relative Motion in Adjacent Orbital Planes}
The satellites in the adjacent orbital planes ${SAT}_{m,n}$ and ${SAT}_{m+1,n}$ have the angular distances in terms of the differential RAAN and mean anomaly as
\begin{equation}\label{Deltau}
    \Delta u = M_{m+1,n} - M_{m,n} = \frac{360}{T} \cdot F \deg
\end{equation}
\begin{equation}\label{DeltaOm}
    \Delta \Omega = \Omega_{m+1,n} - \Omega_{m,n} = \frac{360}{P} \deg
\end{equation}
Since the Walker-Delta constellation satellites are designed to have the same altitude and inclination, the relative motion between ${SAT}_{m+1,n}$ and ${SAT}_{m,n}$ can be described analytically~\cite{JRWertzBook}. The minimum and maximum relative angular distances ($\theta_{min}$ and $\theta_{max}$) are formulated as follows:
\begin{equation}\label{minLambda}
    \sin {\left( \theta_{min} /2 \right)} = \sin{\left( \phi_{R} /2 \right)} \cos{\left( i_{R} /2 \right)}
\end{equation}
\begin{equation}\label{maxLambda}
    \cos{\left( \theta_{max} /2 \right)} = \cos{\left( \phi_{R} /2 \right)} \cos{\left( i_{R} /2 \right)}
\end{equation}
where $i_{R}$ is the relative inclination and $\phi_{R}$ is the relative phase. 

\begin{figure}[htb!]
	\centering\includegraphics[width=3.5in]{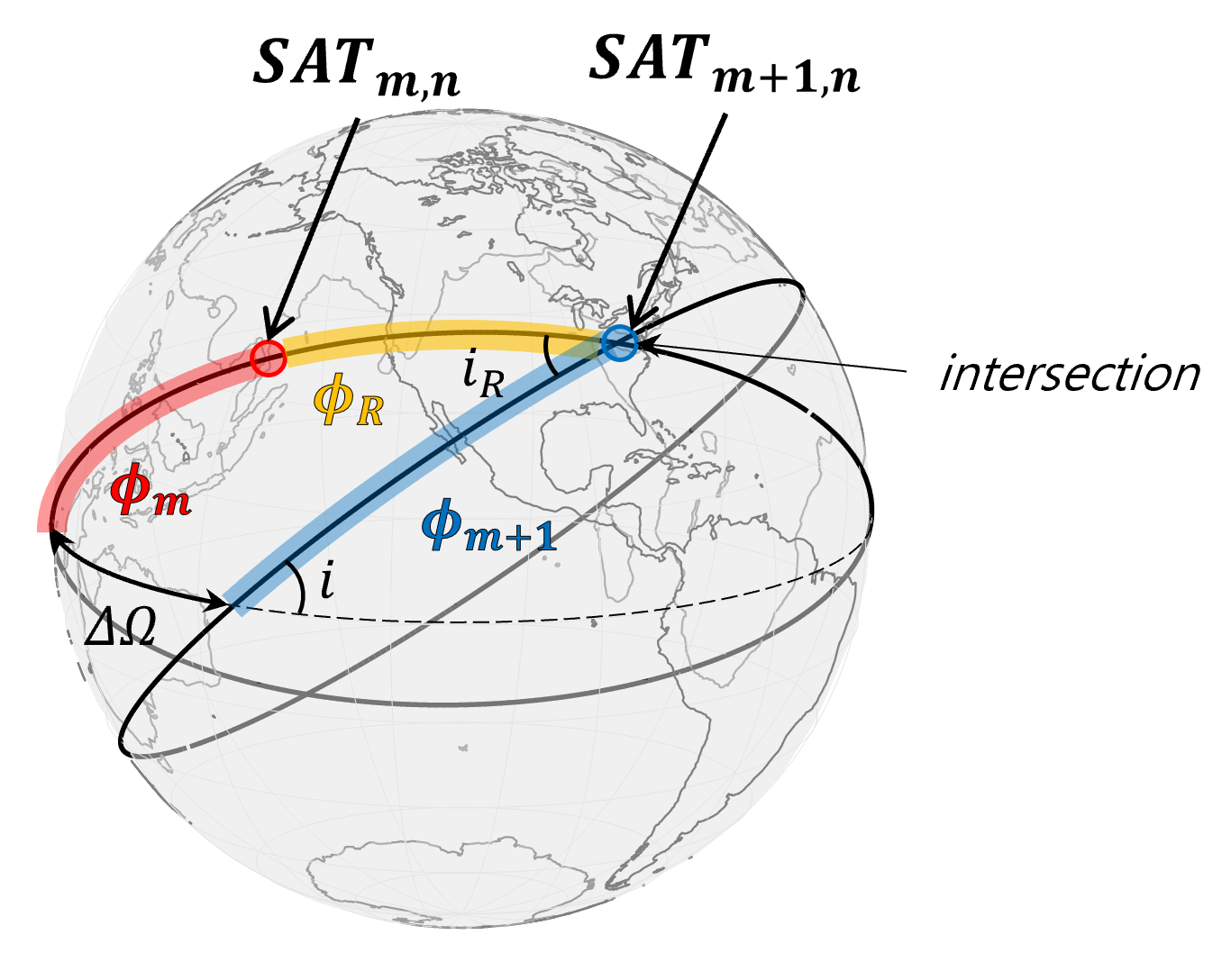}
	\caption{Relative motion of the satellites in the adjacent orbital planes}
	\label{figRelMotion}
\end{figure}

The relative inclination $i_{R}$ in Figure~\ref{figRelMotion} is the angle between the orbital planes measured at the orbital intersection and is derived from spherical trigonometry as
\begin{equation}\label{iR}
\begin{aligned}
    &\cos{i_{R}} = \cos{i}^{2} + \sin{i}^{2} \cos{\Delta \Omega} \\
    & \rightarrow i_{R} = i_{R} \left(i; P \right)
\end{aligned}
\end{equation}
where Eq.~\eqref{DeltaOm} is used. 

The spherical triangle in Figure~\ref{figRelMotion} derives the geometric relationship between the relative phase $\phi_{R}$ and the angular distances $\phi_{m}$ and $\phi_{m+1}$ as follows:
\begin{equation}\label{relPhase}
    \phi_{m} + \phi_{R} - \phi_{m+1} = 180 - 2 \phi_{m+1}
\end{equation}
where the spherical trigonometric rule that the differential arc between the intersected orbits is $180 - 2 \phi_{m+1}$ is used. 
The relative phase $\phi_{R}$ is obtained by reorganizing Eq.~\eqref{relPhase}
\begin{equation}\label{phiR1}
\begin{aligned}
    \phi_{R} &= 180 - 2 \phi_{m+1} + \left( \phi_{m+1} - \phi_{m} \right) \\
    & = 180 - 2 \phi_{m+1} + \Delta u
\end{aligned}
\end{equation}
where the differential angular distance $\phi_{m+1} - \phi_{m}$ is the relative argument of the latitude $\Delta u$ in the Walker-Delta constellation. 
The formula to calculate $\phi_{m+1}$ is 
\begin{equation}\label{phimp1}
\begin{aligned}
    &\tan{\phi_{m+1}} = \frac{\tan{\left( 90 - \Delta \Omega /2 \right)}}{\cos{i}} \\
    & \rightarrow \phi_{m+1} = \phi_{m+1} \left( i; P \right)
\end{aligned}
\end{equation}
As a result, Eq.~\eqref{phimp1} provides the argument of $\phi_{R}$ as follows:
\begin{equation}
    \phi_{R} = \phi_{R} \left( i; T,P,F \right)
\end{equation}

The inter-satellite link (ISL) constrains the range of relative motion so that the signal is not interfered within the link margin. Therefore, the minimum and maximum relative distances in Eqs.~\eqref{minLambda} and \eqref{maxLambda} within the specified range guarantee a smooth ISL communication. 

The relative motion of the adjacent orbital plane is described in Eqs.~\eqref{iR}, \eqref{phiR1}, and~\eqref{phimp1} and explains the relative motion between ${SAT}_{m,n}$ and ${SAT}_{m+1,n}$ where $m=1,...,P-1$ and $n=1,...,S$. When $m=P$, the relative motion between ${SAT}_{P,n}$ and ${SAT}_{1,n+F}$ is proven to be the same as the one between ${SAT}_{m,n}$ and ${SAT}_{m+1,n}$ by Eqs.~\eqref{OmMWD1}, \eqref{OmMWD2}, \eqref{iR}, \eqref{phiR1} and~\eqref{phimp1}. As a result, if Eqs.~\eqref{minLambda} and~\eqref{maxLambda} satisfy the constraint on the ISL link, then all ISL links are connected without any isolated links.

\section{Simulation Results}
\subsection{Continuous Coverage Analysis}
The constellation is designed using the three constellation design methods introduced in the previous sections: Walker-Delta constellation, quasi-symmetric CGT, and BILP CGT constellations. The seed satellites for the three constellations are designed to have the repetition period of $\nu = 14/1$. The inclination is 42 $\deg$ which is 3 $\deg$ --5 $\deg$ higher than the area of interest\cite{JRWertzPaper,XFu}. The minimum elevation angle $\varepsilon_{min}$ is 15 $\deg$ and the target point is located in Seoul. The Walker-Delta constellation is analyzed assuming twobody motion, and the target area is a circle with a radius of 100 km around Seoul. The CGT constellations assume $J_{2}$ perturbation and a single target point. The sampling time or time step $t_{step}$ is 1 and 300 s for the Walker-Delta and CGT constellations, respectively, and the simulation time horizon is 1 day for both. 

\subsubsection{Walker-Delta Constellation}

\begin{figure}[t!]
	\centering\includegraphics[width=6in]{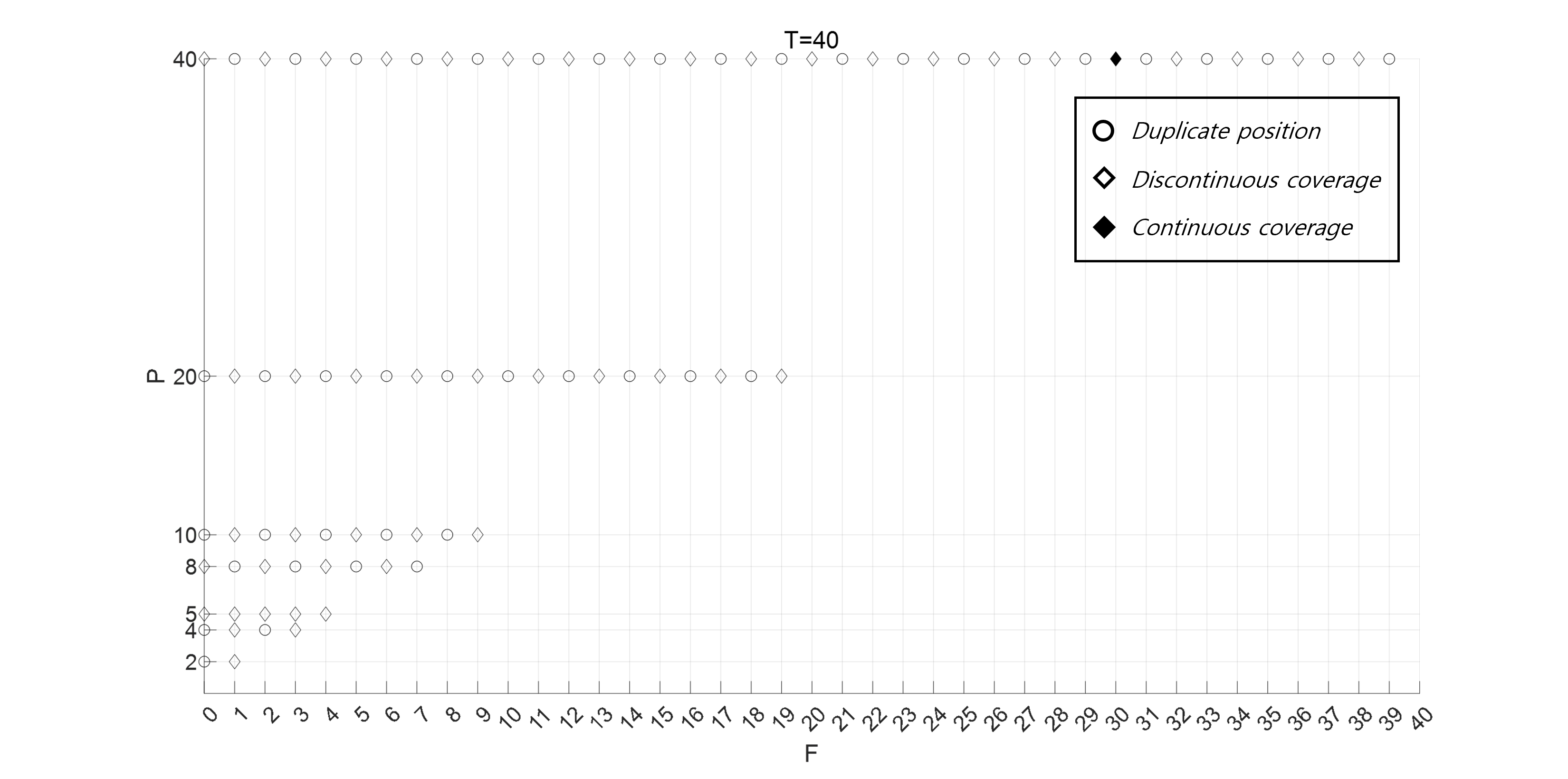}
	\caption{Bounded Voronoi diagram result of Walker-Delta constellation ($i$: $T$ = 42: 40)}
	\label{figSimWD1}
\end{figure}

Figure~\ref{figSimWD1} depicts the bounded Voronoi diagram simulation result of the Walker-Delta constellation. The global search of a smaller number of satellites obtains that 40 satellites at an inclination of 42 $\deg$ is the minimum number of satellites required to achieve the continuous coverage using the Walker-Delta constellation. The empty circles represent the $T/P/F$ parameters with duplicate positions in Eq.~\eqref{dupWD} and are precluded from the coverage analysis. The empty diamond markers indicate that the parameters did not achieve the continuous coverage. The continuous coverage solution is $i$: $T/P(S)/F$ = $42$: $40/40(1)/30$ and denoted as the black diamond. The ECA of this solution ($\lambda^{*}$) is $15.86 \deg$.

\subsubsection{Common Ground-track Constellation}

\begin{figure}[htb!]
	\centering\includegraphics[width=3.5in]{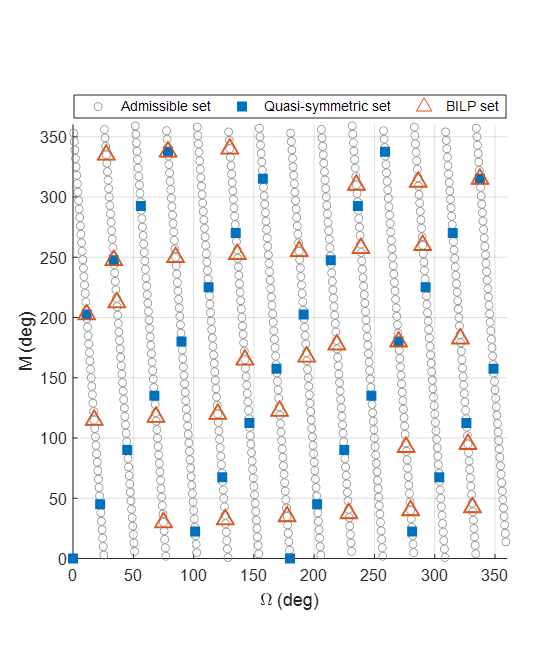}
	\caption{Configurations of quasi-symmetric and BILP CGT constellations in ($\Omega, M$) space}
	\label{figSimCGT1}
\end{figure}

\begin{figure}[htb!]
	\centering\includegraphics[width=6in]{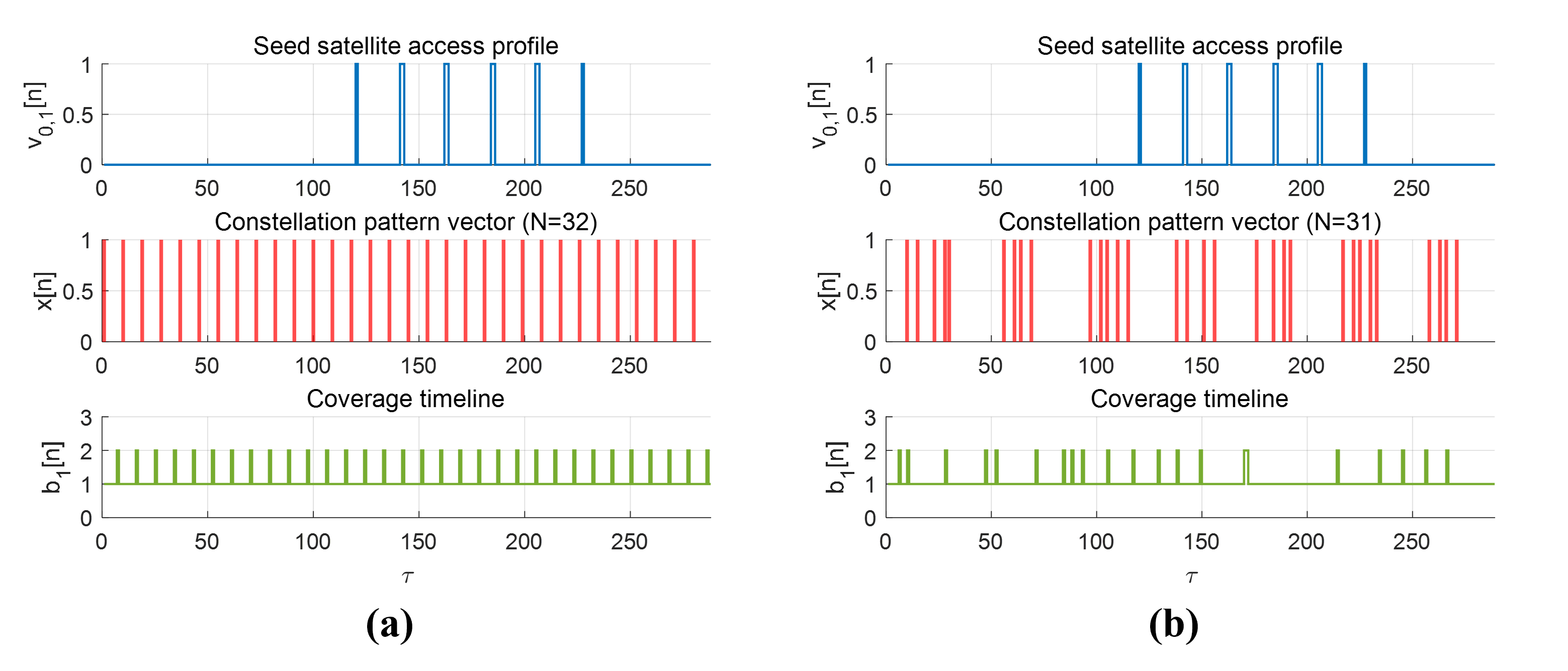}
	\caption{APC Decomposition of (a) quasi-symmetric and (b) BILP CGT constellation}
	\label{figSimCGT2}
\end{figure}

The quasi-symmetric constellation pattern vector $x_{qs}$ and BILP constellation pattern vector $x_{bilp}$ are obtained as 
\begin{equation}\label{xqs}
    x_{qs} = \begin{cases}
        1 & \text{for } n = \{0,9,18,27,36,45,54,63,72,81,90,99,108,117,126,135,... \\
          & 144,153,162,171,180,189,198,207,216,225,234,243,252,261,270,279\} \\
        0 & \text{otherwise}
    \end{cases}
\end{equation}
\begin{equation}\label{xbilp}
    x_{bilp} = \begin{cases}
        1 & \text{for } n = \{9,14,22,27,29,55,60,63,68,96,101,104,109,114,137,... \\
          & 142,150,155,175,183,188,191,216,221,224,229,232,257,262,265,270\} \\
        0 & \text{otherwise}
    \end{cases}
\end{equation}

The CGT constellation's pattern reveals its characteristics in the $\left( \Omega, M \right)$ space, such as period ratio and symmetricity (Figure~\ref{figSimCGT1}). The gradient of the admissible set is $-N_{P}/N_{D}$ and is equal to $-\nu$ by Eqs.~\eqref{CGTomM} and \eqref{tauNpNd}. The constellation pattern vectors are laid on the points along the admissible set. The quasi-symmetric set configures symmetrically in Figure~\ref{figSimCGT1} as $\textbf{\textit{x}}_{qs}$ in the second panel of Figure~\ref{figSimCGT2}a is equally spaced. Because $N_{qs}$ is 32 and $L$ is 288, $L/N_{qs}$ is divided into an integer $9$, and the spacing is perfectly symmetric. On the other hand, the BILP constellation pattern vector is irregularly spaced in Figures~\ref{figSimCGT1} and~\ref{figSimCGT2}b. However, the BILP constellation achieves the smaller number of satellites $N_{bilp}$ as 31, while both constellations exhibit the single-fold coverage.

\subsection{Inter-satellite Link Continuity Analysis}

\subsubsection{Walker-Delta Constellation}

\begin{figure}[t!]
	\centering\includegraphics[width=6in]{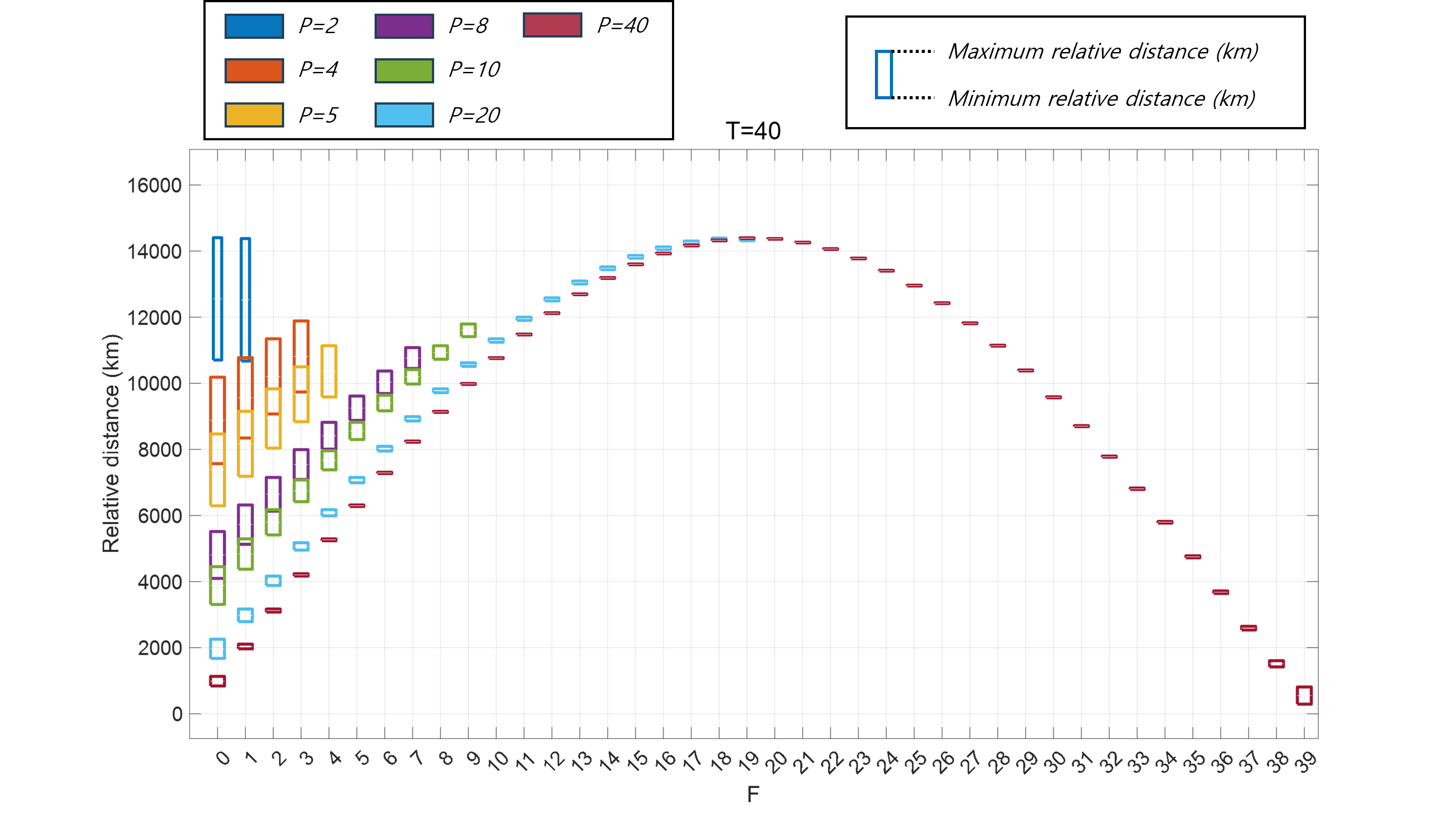}
	\caption{Relative motion between the adjacent orbital plane of Walker-Delta constellation ($i$: $T$ = 42: 40)}
	\label{figSimWD3}
\end{figure}

The minimum and maximum relative distances are the line distances calculated from the angular distances in Eqs.~\eqref{minLambda} and \eqref{maxLambda} (Figure~\ref{figSimWD3}). The upper and lower bounds of the bars indicate the relative distance ranges in the adjacent orbital planes. The colors of the bars distinguish the number of planes, and the $x$ axis represents the $F$ numbers. As this graph shows the relative motion at a glance, it is a useful tool for ISL connectivity analysis. The continuous coverage solution $i$: $T/P(S)/F$ = $42$: $40/40(1)/30$ has a relative motion range from $9559.77$ km to $9589.64$ km.

\subsubsection{Common Ground-track Constellation}

\begin{figure}[b!]
	\centering\includegraphics[width=6in]{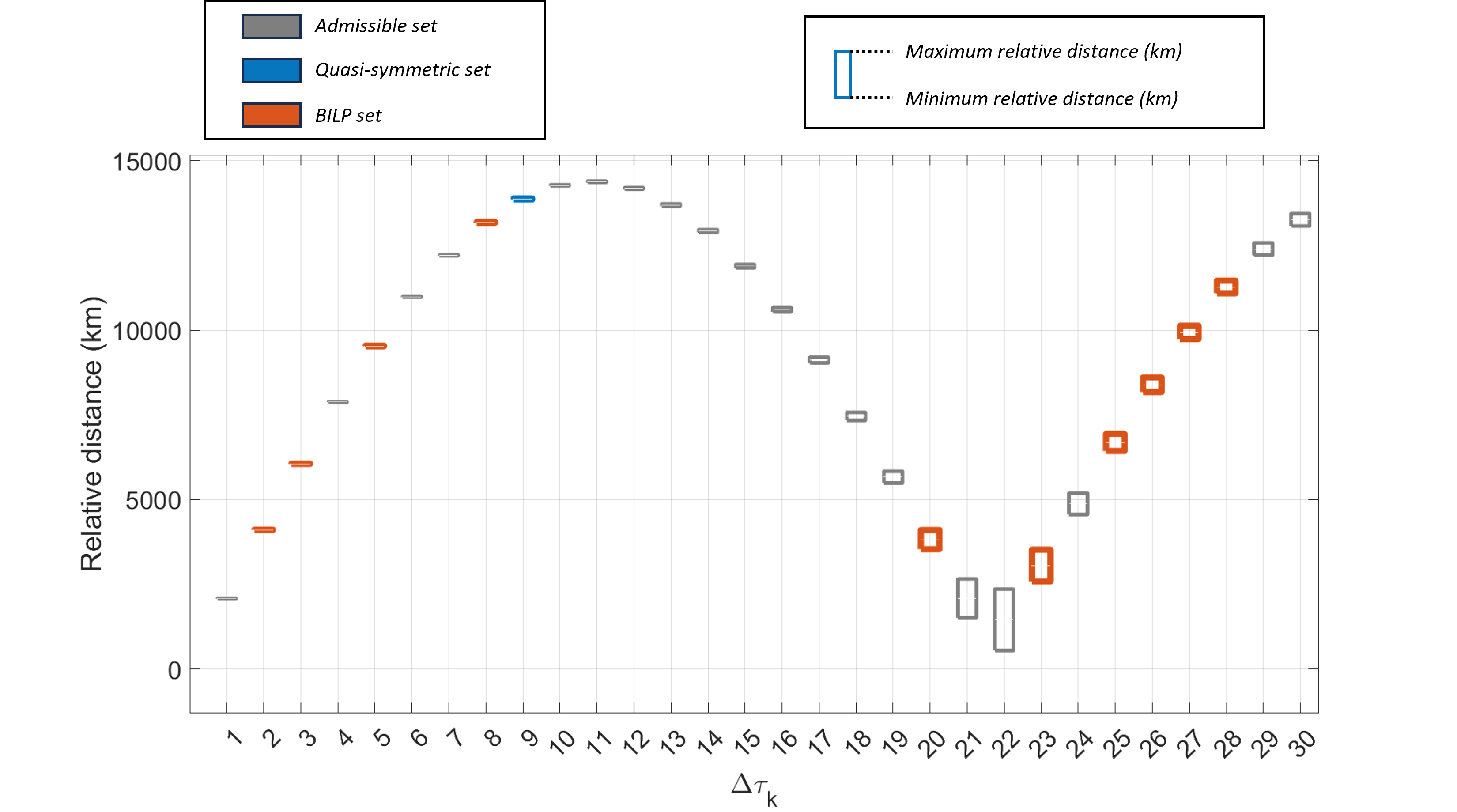}
	\caption{Minimum and maximum relative distance between the adjacent orbital plane of quasi-symmetric and BILP CGT constellations}
	\label{figSimCGT3}
\end{figure}

The relative motion equations, Eqs.~\eqref{iR}, \eqref{phiR1}, and~\eqref{phimp1}, imply that the relative motion of two orbits with the same altitude and inclination is a function of the inclination, relative RAAN, and relative argument of latitude. Let us define a set of temporal location $\tau_{k}$ in Eq.~\eqref{constPV} as $\boldsymbol{\tau}_{k}$. Then, $\boldsymbol{\tau}_{k,qs}$ and $\boldsymbol{\tau}_{k,bilp}$ are calculated as Eqs.~\eqref{xqs} and~\eqref{xbilp}. The temporal location $\tau_{k}$ derives the RAAN $\Omega_{k}$ as shown in Eq.~\eqref{OmkCGT}\cite{HLee2020}.
\begin{equation}\label{OmkCGT}
    \Omega_{k} = \tau_{k} \cdot \frac{2\pi N_{D}}{L} + \Omega_{0}
\end{equation}
where the subscript `0' means that the variable is relevant to the seed satellite. Equation~\eqref{CGTomM} describes the relationship between $\Omega_{k}$ and $M_{k}$. 

Let us define $\Delta \tau_{k}$ as the difference between consecutive $\tau_{k}$ values:
\begin{equation}\label{dnk}
    \Delta \tau_{k} = \begin{cases}
        \tau_{k+1} - \tau_{k} & \text{for $k = 1,...,T-1$} \\
        \tau_{k} + L - \tau_{1} & \text{for $k = T$}
    \end{cases}
\end{equation}
Thus, Eqs.~\eqref{xqs}, \eqref{xbilp}, and~\eqref{dnk} are used to calculate $\Delta \tau_{k}$ for the two CGT constellations as
\begin{equation}
    \Delta \tau_{k,qs} = 9
\end{equation}
\begin{equation}
    \Delta \tau_{k,bilp} = 2, 3, 5, 8, 20, 23, 25, 26, 27, 28
\end{equation}
Equation~\eqref{OmkCGT} derives the relative RAAN $\Delta \Omega_{k}$ and the relative mean anomaly $\Delta M_{k}$ as 
\begin{equation}
\begin{cases}
    \Delta \Omega_{k,qs} = 11.25 \deg \\
    \Delta M_{k,qs} = 202.50 \deg    
\end{cases}
\end{equation}
\begin{equation}
\begin{cases}
    \Delta \Omega_{k,bilp} = 2.5, 3.75, 6.25, 10.00, 25.00, 28.75, 31.25, 32.50, 33.75, 35.00 \deg \\
    \Delta M_{k,bilp} = 10.00, 220.00, 230.00, 247.50, 265.00, 272.50, 282.50, 307.50, ...\\
    \indent\indent\indent\indent\indent 317.50, 325.00 \deg
\end{cases}
\end{equation}

The minimum and maximum relative distances in Eqs.~\eqref{minLambda} and \eqref{maxLambda} are depicted in Figure~\ref{figSimCGT3}. The minimum and maximum relative distances appear almost identical for some cases because the differences are less than 1000 km. The minimum and maximum relative distances for the quasi-symmetric CGT constellation are 13854.32 and 13886.49 km, respectively. The BILP CGT constellation has various values of $\Delta \tau_{k}$. When $\Delta \tau_{k}$ is 8, the minimum and maximum relative distances reach their largest values at 13164.56 and 13191.33 km, respectively.

\section{Conclusion}

This paper investigates the continuous coverage analysis methods and the inter-satellite link connectivity analysis method for communication satellite constellations. The bounded Voronoi diagram is used to design a homogeneous constellation that ensures continuous regional and global coverage. The APC decomposition,  based on the grid method, can be implemented for the CGT constellation's coverage analysis. The relative motion in adjacent orbital planes yields the analytical solutions that must be constrained within the inter-satellite link range. 

The coverage performance of the Walker-Delta constellation was analyzed using the bounded Voronoi diagram as an example. Two types of CGT constellations, the quasi-symmetric and the BILP, were analyzed using APC decomposition to compare their coverage performance. The BILP optimal CGT constellation has an asymmetric configuration but achieves a smaller number of satellites. However, the relative motion range of the Walker-Delta constellation is shorter and more consistent, which implies that the Walker-Delta constellation has advantages for inter-satellite links. The relative motion of the BILP constellation has a variety of ranges due to its asymmetry, but satellites are located closer than the quasi-symmetric constellation. In summary, the BILP constellation is advantageous in terms of the number of satellites required for a single-fold coverage. However, the Walker-Delta constellation may have a shorter and more stable relative motion range, which is beneficial for inter-satellite links.

\section{Acknowledgment}
This  work was supported by the Korea Research Institute for defense Technology Planning and Advancement (KRIT) grant funded by the Korean government (DAPA (Defense Acquisition Program Administration)) (KRIT-CT-22-040, Heterogeneous Satellite Constellation-based ISR Research Center, 2024).

\newpage

\section{Notation}
\begin{tabular}{r l}
	$a$ & semi-major axis \\
	$\textbf{\textit{b}}$ & coverage timeline \\
    $\delta$ & Kronecker delta function\\
        $e$ & eccentricity \\
        $\textbf{\textit{f}}$ & coverage requirement vector \\
        $i$ & inclination \\
        $\textbf{\textit{f}}$ & coverage requirement \\

        $h$ & altitude \\
        $j$ & index for grid point \\
        $J_{2}$ & coefficient for J2 perturbation \\
        $\mathcal{J}$ & set of grid points \\
        $k$ & index for satellites \\
        $L$ & number of discrete times (length of discrete time variable) \\
        $m$ & index for orbital planes \\
        $n$ & index for satellites on a plane \\

        $N_{P}$ & revolutions to repeat \\
        $N_{D}$ & days to repeat \\
        $P$ & number of orbital planes \\
        $F$ & Phasing parameter \\
        $R_{E}$ & Earth radius \\
        $p$ & semilatus rectum \\
        $\mu_{E}$ & standard gravitational parameter of Earth \\
        $S$ & number of satellites on an orbital plane \\
        $T$ & total number of satellites \\
        $T_{r}$ & repetition period of RGT orbit \\
        $T_{S}$ & nodal period of the satellite \\
        $T_{G}$ & nodal period of greenwich \\
        $T_{sim}$ & simulation time \\
        $t_{step}$ & time step \\
        $t$ & continuous time variable \\
        $t_{0}$ & epoch time \\
        $t_{F}$, $t_{P}$ & Walker-Delta pattern repetition period\\
        $u$ & argument of latitude \\
        $\textbf{\textit{v}}$ & access profile \\
        $\textbf{V}$ & access profile circulant matrix \\
        \textbf{\textit{x}} & constellation pattern vector \\
        $\mathbb{Z}_{2}$ & binary integer number set \\

        $\epsilon$ & elevation angle \\
        $\eta$ & angular size of payload's coverage \\
        $\rho$ & angular radius of Earth \\
        $\lambda$ & Earth central angle \\
        $\nu$ & period ratio \\
        $\phi$ & phase angle \\
        $\tau$ & discretized time variable \\
        $\omega$ & argument of perigee \\
        $\omega_{E}$ & Earth rotation speed \\
        $\omega_{orb}$ & orbital angular speed \\
        $\Omega$ & right ascension of ascending node \\
        $\xi$ & spacing constant \\
        $\theta$ & angular distance \\
        
\end{tabular} \\


\bibliographystyle{AAS_publication}   

\end{document}